\documentclass[12pt]{iopart}

\usepackage{iopams,graphicx} 

\newcommand{\gs}{\gamma_S}

\newcommand{\ssbar}{{\rm s}\bar{\rm s}}
\newcommand{\uubarp}{\langle {\rm u}\bar{\rm u} \rangle}
\newcommand{\ddbarp}{\langle {\rm d}\bar{\rm d} \rangle}
\newcommand{\ssbarp}{\langle {\rm s}\bar{\rm s} \rangle}

\begin{document}

\title[]{Strangeness production from SPS to LHC}

\author{F Becattini*}
\address{Universit\`a di Firenze and INFN Sezione di Firenze,
Via G. Sansone 1, I-50019, Sesto F.no, Firenze, Italy}
\ead{becattini@fi.infn.it}
\author{J. Manninen}
\address{INFN Sezione di Firenze,
Via G. Sansone 1, I-50019, Sesto F.no, Firenze, Italy}

\begin{abstract}

Global strangeness production in relativistic heavy ion collisions at SPS
and RHIC is reviewed. Special emphasis is put on the comparison with the 
statistical model and the canonical suppression mechanism. It is shown that 
recent RHIC data on strange particle production as a function of centrality 
can be explained by a superposition of a fully equilibrated hadron gas and 
particle emission from single independent nucleon-nucleon collisions in the 
outer corona. 

\end{abstract}


\section{Introduction}

The enhancement of relative (to u,d quarks) strange quark production in high energy 
heavy ion collisions with respect to elementary collisions has been predicted long
time ago to be a signature of the Quark Gluon Plasma (QGP) formation \cite{rafmu}. 
The idea was that chiral symmetry restoration favours strange quark production 
because of the reduced mass compared to its zero temperature constituent value.
This abundant strangeness production could be observed provided that it survives 
hadronization, i.e. if the early produced strange quarks coalesce into hadrons 
without reannihilating. A specific prediction of such a mechanism is the enhancement 
of multiply strange particles, especially hyperons.

\begin{figure}
\begin{center}
  \includegraphics[height=.35\textheight]{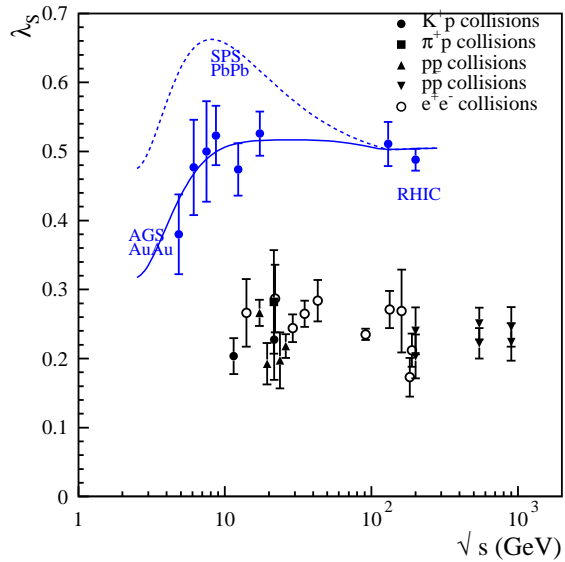}
  \caption{Wroblewski ratio in elementary and heavy ion collisions as estimated
   from statistical model fits. The superimposed dashed line is the predicted 
   value from a fully equilibrated hadron gas, the solid line the interpolation
   from hadron gas with extra strangeness suppression $\gs$.}
\end{center}
\label{ls}
\end{figure}

These phenomena have been indeed observed: the ratio of newly produced strange
to u,d quarks (the so-called Wroblewski ratio $\lambda_S = \ssbarp/2(\uubarp + \ddbarp)$
shows about a factor 2 increase going from elementary to heavy ion collisions,
(see fig.~\ref{ls}) as first observed in ref.~\cite{bgs}, and the hyperons shows 
a clear hyerarchical enhancement in central Pb-Pb collisions with respect to 
peripheral Pb-Pb and p-Pb collisions at top SPS energy ($\sqrt s_{NN} = 17.2$ GeV), 
as observed by the WA97-NA57 collaboration \cite{wa97}. Also, it seems that this ratio
increases quickly in heavy ion collisions as a function of centre-of-mass energy
going from 1 to few GeVs and stay constant thereafter. 

The big question is what is the origin of this observed strangeness enhancement.
Is the original prediction of generation in the plasma and subsequent coalescence 
still viable? Or, rather, are the excess strange quarks produced essentially at 
hadronization? Or, finally, is strangeness produced during an intense hadronic 
re-scattering stage, according to transport models {\it ansatz}?
Before trying to answer these questions, it is necessary to address a preliminary
very important issue, i.e. whether we have produced a completely equilibrated 
hadron gas or not. If we have a completely equilibrated hadron gas, strangeness
content is completely determined and gives information on freeze-out state, but 
it is not a probe of earlier stage of the process. Solving this problem may have 
a considerable impact on our understanding of strangeness production in relativistic 
heavy ion collisions.

\section{Statistical model and strangeness undersaturation}

The main tool to probe the formation of an equilibrated hadron gas are the fits
of the measured particle multiplicities or ratios to the statistical model, that
is the ideal hadron-resonance gas. Many authors have performed such analyses 
trying to pinpoint the thermodynamical parameters of the hadron emitting source
at the chemical freeze-out and their conclusions are vastly different in this 
respect. Some \cite{pbm} conclude that a completely equilibrated hadron gas has been
produced throughout the examined centre-of-mass energy range (from low AGS to
RHIC), others \cite{raf} that this never occurs. The reason of such a dramatic
difference in physical conclusions resides on one hand on data selection and, 
on the other hand, on parameter choice in fitting procedure.
\begin{figure}[h]
\begin{minipage}[t]{7.cm}
\includegraphics[height=.3\textheight]{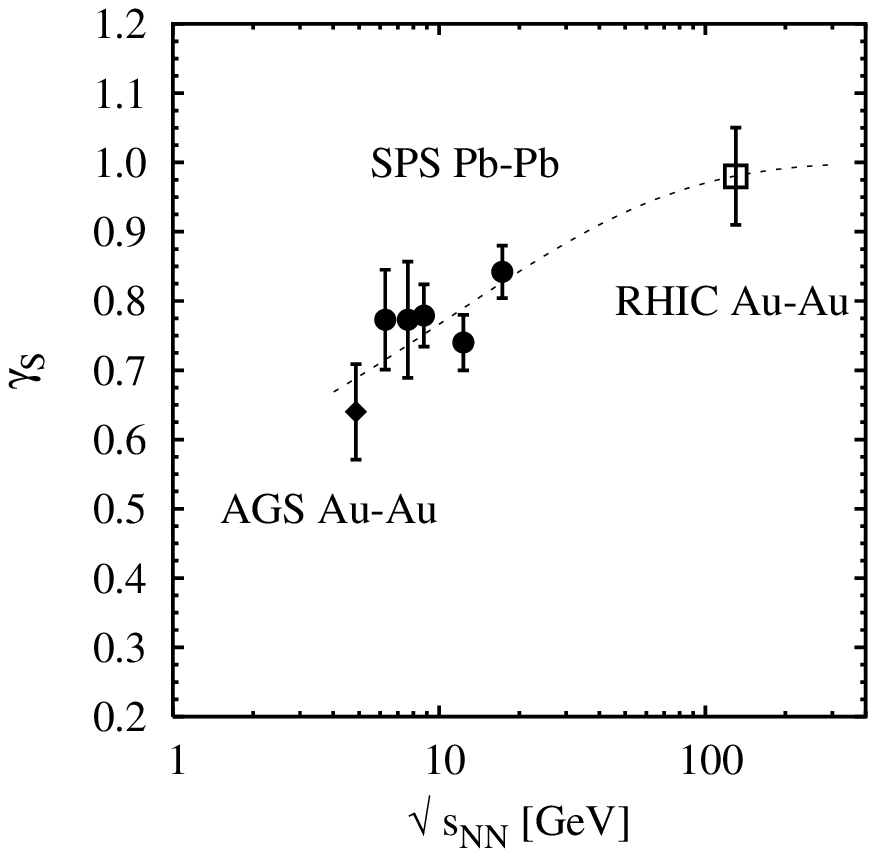}
\end{minipage}
\hskip1cm
\begin{minipage}[t]{7.cm}
\includegraphics[height=.3\textheight]{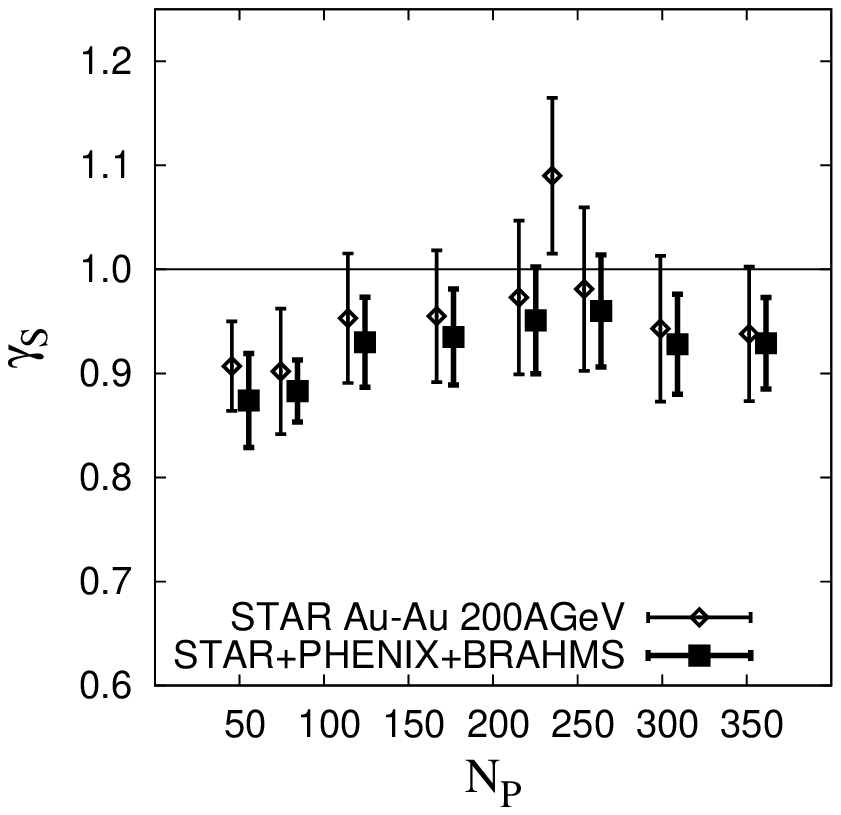}
\end{minipage}
\caption{Left panel: $\gs$ as a function of centre-of-mass energy in central 
 heavy ion collisions (from ref.~\cite{becahi4}). Right panel: $\gs$ as a 
 function of centrality in Au-Au collisions at $\sqrt s_{NN} = 200$ GeV in 
 central heavy ion collisions \cite{prep}.}
\label{gs}
\end{figure}

The conclusion that a completely equilibrated hadron gas is found relies on the
use of midrapidity densities as experimental input compared with integrated 
yields of a single fireball at full hadrochemical equilibrium as theoretical model. 
The underlying idea is that, being the QGP fireball expected at midrapidity, 
its properties can be probed by using midrapidity densities. However, such an 
approach requires the existence of a region around midrapidity (a plateau)
where the thermodynamical parameters do not vary much. Conversely, 
if the observed particle rapidity distributions are not sufficiently wider than 
that of a single fireball at the thermal freeze-out temperature, the use of 
midrapidity densities instead of integrated multiplicities artificially enhances 
heavier particles which have, in general, narrower rapidity width than lighter 
particles. This has two biasing effects: 
increasing the estimated temperature and enhancing the yield of particles carrying
strange quarks which are generally heavier than non-strange ones \cite{becahi3}
so that the strangeness undersaturation parameter $\gs$ turns out to be 
approximately 1 in these fits and essentially unnecessary. 

In fact, up to SPS energies, the rapidity distributions are not wide enough to allow
the use of midrapidity densities. For instance, the pion rapidity distribution   
at $\sqrt s_{NN} = 17.2$ GeV has a dispersion width of 1.3, while that of a 
single fireball at rest with $T=125$ MeV (the thermal freeze-out temperature)
is about 0.8, hence not much smaller. Conversely, the measured width at RHIC
at $\sqrt s_{NN} = 200$ GeV is 2.1, which is reasonably larger than 0.8. 
Thus, the use of midrapidity densities allows to determine the thermodynamical
parameters of the average fireball at midrapidity at RHIC energies onwards
(roughly from 100 GeV), but not at SPS and lower energies. Therein, fits to
full phase space yields provide a more appropriate, though amendable, estimate 
of the chemical freeze-out parameters. For recent studies including rapidity-
dependent chemical potentials see refs.~\cite{bronio,becacley}.

As a consequence, a strangeness undersaturation parameter $\gs < 1$ is needed to 
describe particle multiplicities in central heavy ion collisions. This parameter
shows an increasing trend from AGS to RHIC, where it attains its maximal value 1 
(see fig.~\ref{gs}). Moreover, a $\gs <1$ is also needed in peripheral collisions 
at RHIC for midrapidity densities, as shown in fig.~\ref{gs}. 

\section{Canonical suppression}

It has been argued \cite{redlich} that the observed strangeness undersaturation 
(i.e. $\gs < 1$) at energies lower than SPS is owing to the so-called canonical 
suppression effect. Namely, strange particles are further suppressed in pp 
collisions and peripheral relativistic heavy ion collisions with respect to their 
expected yield in a grand-canonical ensemble (or thermodynamic limit) because 
strangeness is exactly vanishing within a small volume, called strangeness 
correlation volume (SCV), not necessarily coinciding with the global volume. 
Therefore, going from pp collisions to central heavy ion collisions through 
peripheral ones, one expects to observe a relative enhancement of strange particles 
due to approaching the thermodynamic limit, which is hyerarchical: $\Omega$ yield 
increases faster than $\Xi$ which increases faster than $\Lambda$'s or kaons.
Yet, although this hyerarchy of enhancements is observed (see fig.~\ref{phi}), 
neither SPS nor RHIC have observed the saturation which should be there if the SCV 
attains a sufficiently large value. In fact, this means that the SCV only reaches 
its saturation value (the one sufficient for the system to be essentially 
grand-canonical) at RHIC precisely in central collisions, where $\gs \simeq 1$.
This would be quite a striking coincidence. Therefore, we think that canonical 
suppression is quite an unnatural explanation of the data, as already pointed 
out in ref.~\cite{nuxu}.

The best probe to investigate the phenomenon of strangeness undersaturation is
indeed the $\phi$ meson. This is not an open strange particle, thus it is not
canonically suppressed, yet, being a $\ssbar$ state, it must be $\gs^2$ suppressed.
Furthermore, $\phi$ meson has almost no feeding from heavier light-flavoured 
species and its production is entirely direct.

\begin{figure}
\begin{center}
  \includegraphics[height=.4\textheight]{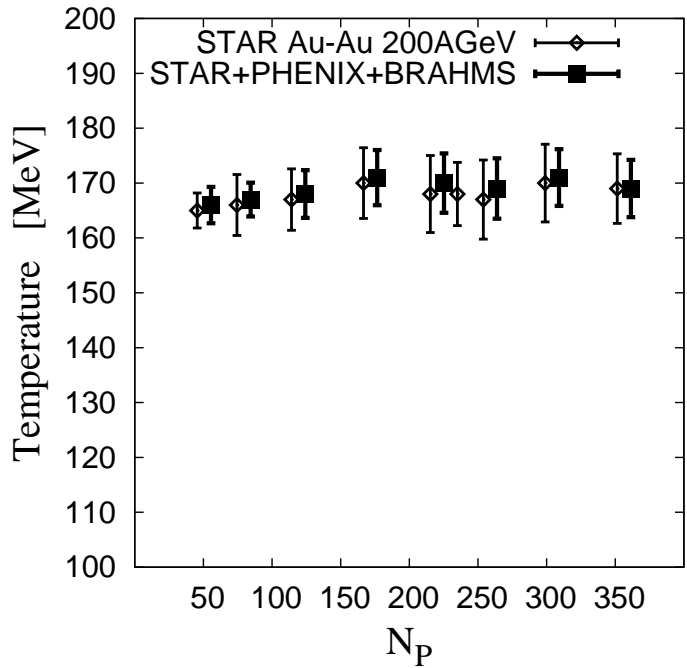}
  \caption{Chemical freeze-out as a function of centrality in Au-Au collisions
  at $\sqrt s_{NN} = 200$ GeV \cite{prep}.}
\end{center}
\label{temp}
\end{figure}

It was pointed out quite early \cite{sollfrank} that a statistical model with canonical 
suppression mechanism, i.e. with SCV as additional parameter, would have not been
able to explain the deviation of the $\phi$ meson yield from its grand-canonical 
value and this has been demonstrated in fits to NA49 multiplicities \cite{becahi3}.
Recently, the STAR collaboration has measured the midrapidity densities of $\phi$
meson very accurately and the observed pattern as a function of centrality clearly
shows (see fig.~\ref{phi}) that these do not scale linearly with the number of
participants, rather the ratio to pp value increases rapidly at very peripheral 
collisions slowly saturating thereafter. This non-linear increase cannot be 
attributed to a variation of the chemical freeze-out temperature because this is 
astonishingly constant as a function of $N_p$ as shown in fig.~\ref{temp} and 
proves that a genuine extra suppression related to the strange quark is needed,
as also reflected in the $\gs$ fitted value (see fig.~\ref{gs}).

\section{Core-corona superposition}

Can we explain $\gs$ in relativistic heavy ion collisions in more fundamantal 
terms? Some years ago \cite{becahi3} R. Stock proposed that 
$\gs < 1 $ in global fits could be the effect of superposing a completely
equilibrated hadron gas ($\gs=1$) originated from the core of the nuclear collision 
(i.e. the hadronization of the plasma) to a corona of single NN collisions where 
particle readily escape the interaction region. Since strangeness is largely 
suppressed in NN collisions with respect to the grand-canonical value while
the temperature is almost the same as we know from pp statistical model analysis 
\cite{becahh,becahi4}, if the number of such single NN collisions accounts for 
a significant fraction of total particle production, a global fit to one 
hadron-resonance gas would actually find $\gs$ significantly less than 1. Indeed, 
this idea proved to be able to satisfactorily reproduce particle multiplicities 
in central C-C, Si-Si and Pb-Pb collisions at top SPS energy. 
\begin{figure}
\begin{center}
  \includegraphics[height=.4\textheight]{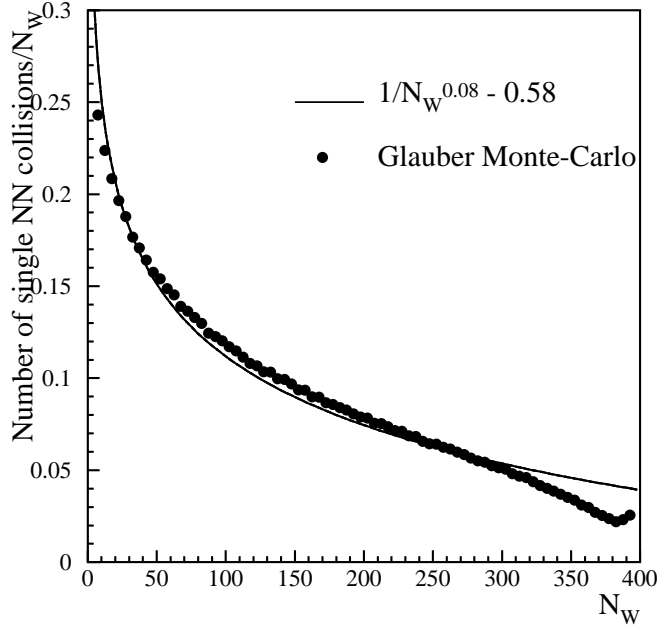}
  \caption{Number of single NN collisions divided by the number of wounded 
  nucleons as a function of number of wounded nucleons according to Glauber 
  Monte-Carlo. The solid line is the interpolation (\ref{glauber}).}
\end{center}
\label{ns}
\end{figure}

This core-corona superposition mechanism has been invoked by Bozek few
years ago \cite{bozek} to reproduce the $K/\pi$ ratio as a function of centrality
in Au-Au collisions and has been recently advocated in a paper 
by Werner \cite{werner} to be capable of explaining some more otherwise 
``mysterious" effects. A sharp superposition of a completely equilibrated 
hadron gas with NN collisions is indeed a zero-order approximation as 
the actual process is certainly more complex with those two extremes continuously 
linked through intermediate steps and indeed in ref.~\cite{werner} a more general
concept of corona has been used, defined as a ``dilute" peripheral region 
distinguished from the ``dense" region in the core. 
Yet, this simple superposition scheme can be a very useful one to understand 
the physics of particle production. Accordingly, the rapidity density at 
midrapidity of any particle species is given by:
\begin{equation}\label{phi1}
 \langle \frac{d n} {d y} \rangle = N_s \langle \frac{d n} {d y} \rangle_{pp} +
f (V_0 - \delta V_0)  \langle \frac{d \rho} {d y} \rangle_{core}
\end{equation}
where $N_s$ is the mean number of single NN collisions, $V_0$ is the initial volume
of the initial nuclear overlapping region, $\delta V_0$ is its thin outer
shell where these single NN collisions occur, $f$ is the growth factor (i.e. how
much this volume expands up to chemical freeze-out) and $d \rho/dy$ is the particle
density per unit rapidity in the core relevant to a completely equilibrated hadron 
gas, i.e. with $\gs=1$. Dividing by the number of wounded nucleons $N_W$ \footnote
{In this work wounded and participant nucleons are synonymous} 
and the rapidity density in pp, we obtain a simple expression from (\ref{phi1}):
\begin{equation}\label{phi2}
  \frac{\langle \frac{d n} {d y} \rangle} {N_W \langle \frac{d n} {d y} \rangle_{pp}}
 \simeq A + \frac{N_s}{N_W}(1-2A)
\end{equation}
where $A$ is an unknown constant. This expression fulfills the constraint that 
both the left and right hand side ought to be $1/2$ when $N_W=2$ and $N_s=1$. 
Remarkably, for the $\phi$ meson the constant $A$ is independent of $N_W$ 
because $T$ is in fact independent of centrality (see fig.~\ref{temp}) 
and $\phi$ does not suffer possible canonical suppression. Indeed, $A$ is the 
asymptotic value of the normalized yield, when the number of participants becomes 
very large; hence, it is normally larger than 1/2 and the second term of the right 
hand side of eq.~(\ref{phi2}) is negative. 

\begin{figure}[h]
\begin{minipage}[t]{7.cm}
\includegraphics[height=.3\textheight]{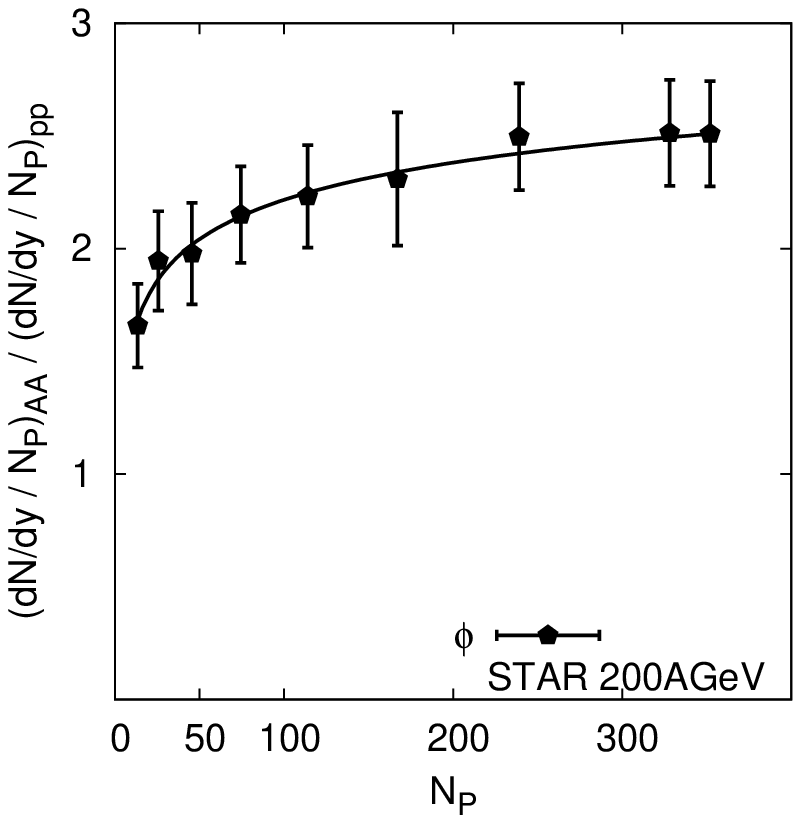}
\end{minipage}
\begin{minipage}[t]{7.cm}
\includegraphics[height=.3\textheight]{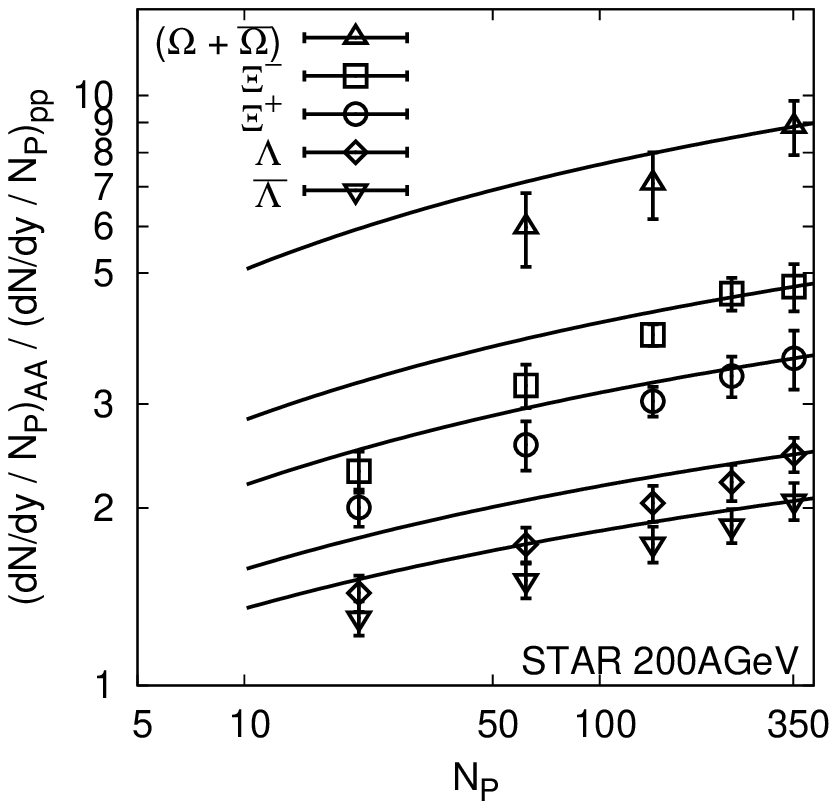}
\end{minipage}
\caption{$\phi$ (left panel) and hyperons (right panel) rapidity density per wounded 
nucleon as a function of participants normalized to pp collisions. Data points 
from STAR \cite{jhchen}; solid lines are the predictions from core-corona 
superposition (see text).}\label{phi}
\end{figure}

The problem now is how to define and estimate the number $N_s$ of single NN collisions.
Ideally, we would like them to be those independent collisions where produced 
particles do not reinteract at all with the surrounding environment. In perfectly 
central collisions, they supposedly are single NN collisions where both
nucleons undergo exactly one collision occurring at the edge of the overlap
region. On the other hand, in extreme peripheral nuclear collisions, they should 
reduce to one NN collision. In all other cases, they 
are tightly related to the NN collisions occurring at the edge of the overlap 
region where only one nucleon from either nucleus is involved, but their number 
cannot be defined in a clearcut way without a full dynamical model of the collision.
However, we can resort to a definition interpolating the perfectly central and 
the extreme peripheral case and relying on the Glauber model. Such a definition 
might be:
\begin{equation}
  N_s \equiv \min[N_{1(a)},N_{1(b)}]
\end{equation}
where $N_{1(a)}$ ($N_{1(b)}$) is the number of nucleons colliding once according
to the Glauber model. We estimated the thus defined $N_s$ as a function of 
centrality by means of a Glauber Monte-Carlo calculation. The resulting $N_s/N_W$ 
ratio can be reasonably fitted for $N_W > 10$ by (see fig.~\ref{ns}): 
\begin{equation}\label{glauber}
 \frac{N_s}{N_W} = \frac{1}{N_W^{0.08}}-0.58
\end{equation}
Plugging (\ref{glauber}) into (\ref{phi2}) we have an expression of the normalized
yield as a function of the number of wounded nucleons, that is centrality, depending
on one unknown parameter $A$. This can be determined by matching the model to the
measured value in the most central bin and then the centrality evolution is 
completely determined. The obtained curve is in impressive agreement with the data, 
as shown in fig.~\ref{phi}; the formula matches the experimental points to a 
high degree of accuracy. This is a clear evidence that the envisaged core-corona 
superposition is able to account for the strangeness undersaturation phenomenon.

The same exercise can be repeated for open strange particles, the result being
shown in fig.~\ref{phi}. It can be seen that the curves match the data 
in the most central bins, while they overestimate the measured points in most
peripheral bins: this is likely due to the canonical suppression effect in 
the core which is not taken into account in the formula (\ref{phi2}) but should 
indeed show up for peripheral collisions. Finally, we observe that other 
definitions of $N_s$ are possible (e.g. $(N_{1(a)}+N_{1(b)})/2$) but they lead 
to similar results.

\section{Discussion and conclusions}

If our interpretation of $\phi$ production as a function of centrality is correct, 
several remarkable consequences are implied. First of all, the enhancement of relative 
strange particle production going from peripheral to central collisions is mainly 
due to a geometrical effect of core-corona superposition. Canonical suppression 
plays a role only in the most peripheral collisions and it is possible that the SCV 
simply coincides with the core volume, what would be a nice fact itself. 
Secondly, the $\phi$ data supports evidence for a completely equilibrated hadron 
gas in the core throughout all centralities at RHIC, whose temperature is constant 
and equal to 165 MeV.

The same conclusion is likely to apply to SPS too. The fact that there 
$\gs \simeq 0.85$ in central collisions \cite{becahi4}, significantly lower than at RHIC,
is related to the lower weight of the core compared to the corona. Indeed, as 
energy decreases, so does the freeze-out volume of the core and the multiplicity 
of particles stemming from it, while the the number of single NN 
collisions decreases only slightly, the NN cross section being slowly varying.
This would nicely explain the mild increase of $\gs$ as a function of centre-of-
mass energy (see fig.~\ref{gs}), nevertheless a complete reanalysis of the data is 
compelling. 
As has been mentioned, early analyses of central collisions at top SPS energy 
based on this picture were fairly succesful \cite{becahi3}, but peripheral 
collisions are indespensable to confirm this idea. In this respect, NA49 is 
going to update preliminary measurements \cite{marek} which were used to determine 
$\gs$ in peripheral Pb-Pb collisions \cite{maiani}.

The constancy of $T$ as a function of centrality (see fig.~\ref{temp}) which
was first observed by the STAR collaboration \cite{barann} is confirmed in our
analysis of RHIC data \cite{prep} to a high degree of accuracy. This stunning 
independence of centrality is hard to reconcile with collisional thermalization, 
as pointed out in ref.~\cite{heinz} as it would require a dramatic dependence
of hadronic reaction rates on temperature. Also, the $\Omega$ yield is very 
difficult to reproduce in such an approach \cite{kapusta} unless invoking the 
existing of massive resonant degrees of freedom \cite{greiner}.  

\section*{Acknowledgments}

We are greatly indebted with H. Caines and J. H. Chen for providing us 
with preliminary STAR measurements of particle production in peripheral Au-Au 
collisions. We warmly thank STAR Collaboration for allowing us to use 
this data. We are grateful to P. Bozek, H. Caines and K. Werner for useful
comments. F. Becattini would like to express its gratitude to the organizers 
of Quark Matter conference.

\section*{References}


\begin{thebibliography}{99}


\bibitem{rafmu}
  J.~Rafelski and B.~Muller, Phys.\ Rev.\ Lett.\  {\bf 48} (1982) 1066

\bibitem{bgs}
  F.~Becattini, M.~Gazdzicki, A.~Keranen, J.~Manninen and R.~Stock,
  Phys.\ Rev.\  C {\bf 69} (2004) 024905

\bibitem{wa97}
 F. Antinori et al., WA97 Collaboration, Phys. Lett. B {\bf 433} (1998) 209. 

\bibitem{pbm}
  A.~Andronic, P.~Braun-Munzinger and J.~Stachel, 
  Nucl.\ Phys.\  A {\bf 772} (2006) 167

\bibitem{raf}
 J.~Rafelski, J.~Letessier and G.~Torrieri,  
  Phys.\ Rev.\  C {\bf 64} (2001) 054907

\bibitem{becahi4}
  F.~Becattini, J.~Manninen and M.~Gazdzicki,
  Phys.\ Rev.\  C {\bf 73} (2006) 044905

\bibitem{prep}
  F. Becattini and J. Manninen, in preparation

\bibitem{becahi3}
  F.~Becattini, M.~Gazdzicki, A.~Keranen, J.~Manninen and R.~Stock,
  Phys.\ Rev.\  C {\bf 69} (2004) 024905

\bibitem{bronio}
  B.~Biedron and W.~Broniowski, Phys.\ Rev.\  C {\bf 75} (2007) 054905

\bibitem{becacley}
  F.~Becattini and J.~Cleymans, J.\ Phys.\ G {\bf 34} (2007) S959

\bibitem{redlich}
  S.~Hamieh, K.~Redlich and A.~Tounsi, Phys.\ Lett.\  B {\bf 486} (2000) 61

\bibitem{nuxu}
  N. Xu, talk given at Critical point and onset of deconfinement, GSI,
  Darmstadt, July 9-13 2007.

\bibitem{sollfrank}
  J.~Sollfrank, F.~Becattini, K.~Redlich and H.~Satz,  
  Nucl.\ Phys.\  A {\bf 638} (1998) 399C

\bibitem{becahh}
  F.~Becattini and G.~Passaleva,
  Eur.\ Phys.\ J.\  C {\bf 23} (2002) 551

\bibitem{bozek}
  P.~Bozek, Acta Phys.\ Polon.\  B {\bf 36} (2005) 3071. 

\bibitem{werner}
  K.~Werner, Phys.\ Rev.\ Lett.\  {\bf 98} (2007) 152301

\bibitem{jhchen}
  J.~H.~Chen for the STAR Collaboration, these proceedings.

\bibitem{marek}
  M. Gazdzicki, private communication.

\bibitem{maiani}
  F.~Becattini, L.~Maiani, F.~Piccinini, A.~D.~Polosa and V.~Riquer,
  Phys.\ Lett.\  B {\bf 632} (2006) 233

\bibitem{barann}
  J.~Adams {\it et al.}  [STAR Collaboration], 
  Phys.\ Rev.\ Lett.\  {\bf 92} (2004) 112301

\bibitem{heinz}
  U.~Heinz and G.~Kestin, ``Universal chemical freeze-out as a 
  phase transition signature,'' PoS C {\bf POD2006}, 038 (2006)
  arXiv:nucl-th/0612105.

\bibitem{kapusta}
  J.~I.~Kapusta, J.\ Phys.\ G {\bf 30} (2004) S351.

\bibitem{greiner}
 J.~Noronha-Hostler, C.~Greiner and I.~A.~Shovkovy, arXiv:0711.0930.

\end{thebibliography}
\end{document}